\documentclass[chi_draft]{sigchi}

\toappear{\scriptsize Permission to make digital or hard copies of all or part of this work for personal or classroom use is granted without fee provided that copies are not made or distributed for profit or commercial advantage and that copies bear this notice and the full citation on the first page. Copyrights for components of this work owned by others than the author(s) must be honored. Abstracting with credit is permitted. To copy otherwise, or republish, to post on servers or to redistribute to lists, requires prior specific permission and/or a fee. Request permissions from permissions@acm.org. \\ {\emph{CHI '20, April 25--30, 2020, Honolulu, HI, USA.} } \\
\copyright~2020 Copyright is held by the owner/author(s). Publication rights licensed to ACM. \\ ACM ISBN 978-1-4503-6708-0/20/04\ ...\$15.00.\\
DOI: https://dx.doi.org/10.1145/3313831.3376625 }
\CopyrightYear{2020}
\setcopyright{acmlicensed}
\usepackage{balance}       % to better equalize the last page
\usepackage{graphics}      % for EPS, load graphicx instead 
\usepackage[T1]{fontenc}   % for umlauts and other diaeresis
\usepackage{txfonts}
\usepackage{mathptmx}
\usepackage[pdflang={en-US},pdftex,breaklinks=true]{hyperref}
\usepackage{color}
\usepackage{booktabs}
\usepackage{textcomp}
\usepackage{multirow}
\usepackage{placeins}
\usepackage{siunitx}
\usepackage{caption}
\usepackage{xcolor}

\usepackage{microtype}        % Improved Tracking and Kerning
\usepackage{ccicons}          % Cite your images correctly!
\usepackage{todonotes}

\makeatletter
\g@addto@macro{\UrlBreaks}{\UrlOrds}
\makeatother

\def\plaintitle{Mirror Ritual: An Affective Interface for Emotional Self-Reflection}

\def\emptyauthor{}
\def\plainkeywords{Authors' choice; of terms; separated; by
  semicolons; include commas, within terms only; this section is required.}

\makeatletter
\def\url@leostyle{%
  \@ifundefined{selectfont}{
    \def\UrlFont{\sf}
  }{
    \def\UrlFont{\small\bf\ttfamily}
  }}
\makeatother
\def\pprw{8.5in}
\def\pprh{11in}

\definecolor{linkColor}{RGB}{6,125,233}
\hypersetup{%
  pdftitle={\plaintitle},
  pdfauthor={\emptyauthor},
  pdfkeywords={\plainkeywords},
  pdfdisplaydoctitle=true, % For Accessibility
  bookmarksnumbered,
  pdfstartview={FitH},
  colorlinks,
  citecolor=black,
  filecolor=black,
  linkcolor=black,
  urlcolor=linkColor,
  breaklinks=true,
  hypertexnames=false
}

\begin{document}

\title{\plaintitle}

\numberofauthors{2}
\author{%
  \alignauthor{Nina Rajcic\\
    \affaddr{SensiLab, Monash University}\\
    \affaddr{Melbourne, Australia}\\
    \email{Nina.Rajcic@monash.edu}}\\
  \alignauthor{Jon McCormack\\
    \affaddr{SensiLab, Monash University}\\
    \affaddr{Melbourne, Australia}\\
    \email{Jon.McCormack@monash.edu}}\\
}

\teaser{ \centering

\includegraphics[width=0.45\textwidth]{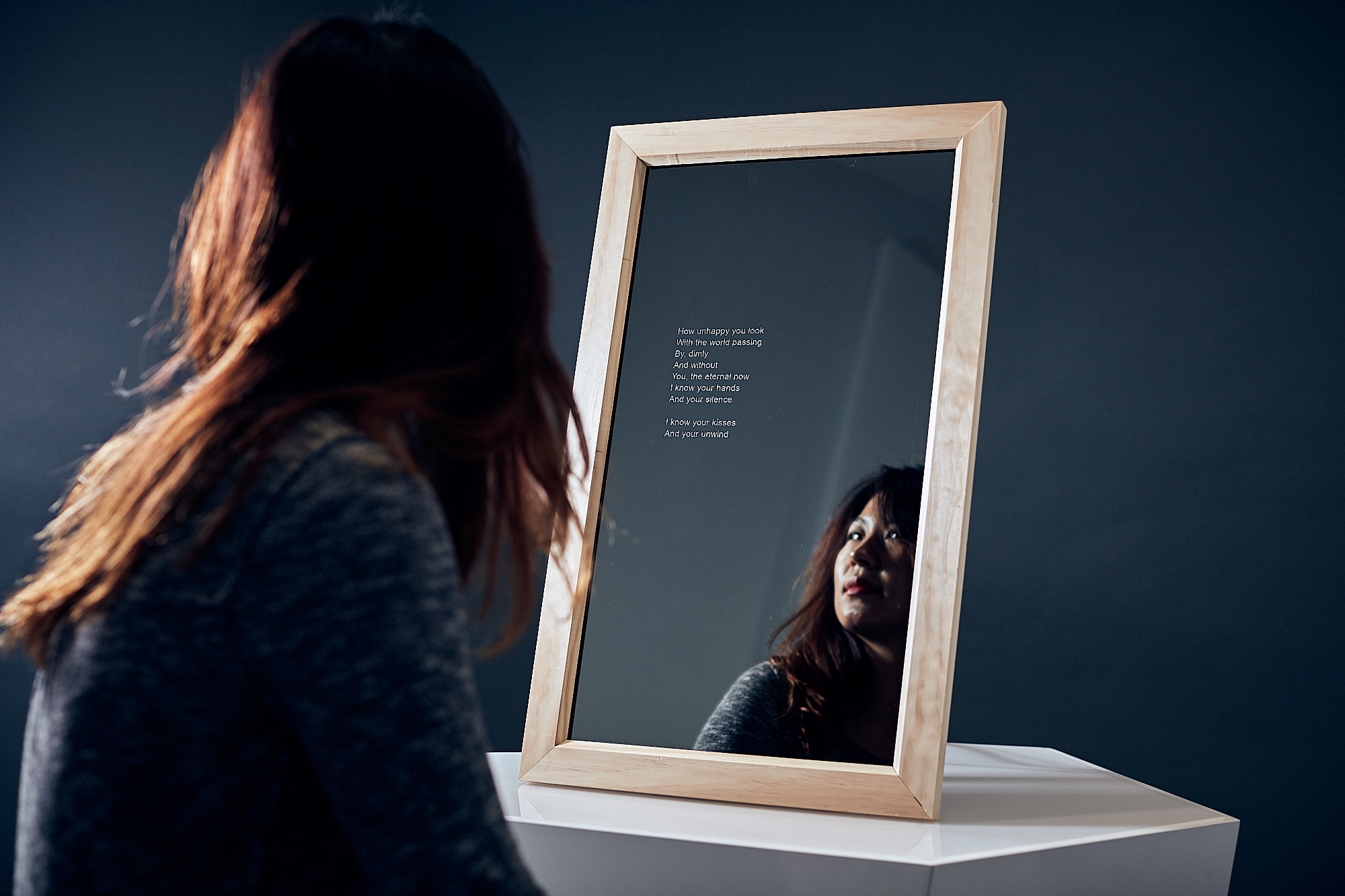} \hspace{10mm}
   \includegraphics[width=0.45\textwidth]{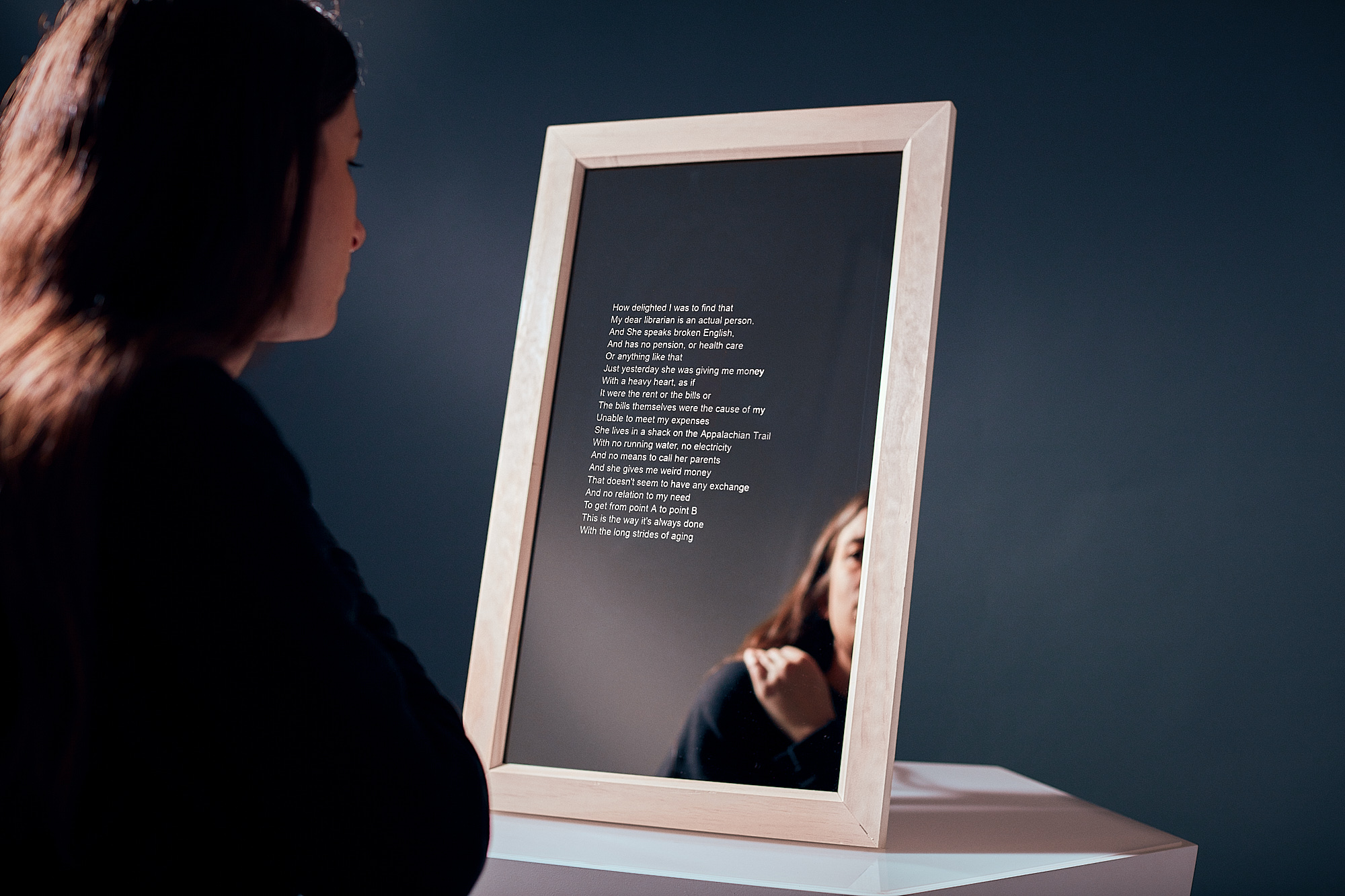}
 \caption{Mirror Ritual: participants interacting with an AI mirror. 
% \Description[A person gazing into a mirror, with poetry displayed on it's surface][]
}
 \label{fig:teaser} }

\maketitle

\begin{abstract}
This paper introduces a new form of real-time affective interface that engages the user in a process of conceptualisation of their emotional state. 
Inspired by Barrett's Theory of Constructed Emotion, `Mirror Ritual' aims to expand upon the user's accessible emotion concepts, and to ultimately provoke emotional reflection and regulation.
The interface uses classified emotions -- obtained through facial expression recognition -- as a basis for dynamically generating poetry. 
The perceived emotion is used to seed a poetry generation system based on OpenAI's GPT-2 model, fine-tuned on a specially curated corpus. 
We evaluate the device's ability to foster a personalised, meaningful experience for individual users over a sustained period. 
A qualitative analysis revealed that participants were able to affectively engage with the mirror, 
with each participant developing a unique interpretation of its poetry in the context of their own emotional landscape.
\end{abstract}

% ACM Classfication



% Author Keywords

% Print the classficiation codes
% Please use the 2012 Classifiers and see this link to embed them in the text: \url{https://dl.acm.org/ccs/ccs_flat.cfm}

\section{Introduction}
\label{s:introduction}

A broad arc of HCI research has shifted over many decades from functional design of virtual and real objects using natural affordances \cite{Norman1990}, through embodied interaction with tangible objects \cite{Dourish2001} to interfaces designed for social and emotional experiences \cite{Wright2004,Vinciarelli_2008,Hassenzahl2013,funology2018}. Coupled with the recent rise in accessible Artificial Intelligence (AI) technologies, new kinds of interactions and interfaces are now being explored \cite{Amershi2019}, in particular those beyond pure utility or function, exploring creative, open-ended interactions and encounters (e.g.~\cite{Koch2019,McCormack2019, Karimi2019, Wallace:2018:SDI:3173574.3173997}). Sustained and meaningful engagement with an AI requires re-conceptualisation of an interface beyond that of a functional tool or pseudo-expressive interactor, because such systems are capable of increasing degrees of autonomy \cite{boden1998,Boden2010} and creative agency \cite{BownM09}. 

This paper introduces a bespoke affective interface that, through sustained engagement, seeks to elicit meaning for its users from their interaction with a non-anthropomorphic AI.  
Inspired by Barrett's \emph{Theory of Constructed Emotion} (TCE), we employ existing AI-based emotion classification techniques, not in an attempt to accurately predict the user's precise emotional state, but instead to engage the user in the conceptualization of their feelings and experiences.

We appropriate an everyday object -- the mirror -- augmenting it with machine vision and artificial intelligence to allow for both literal and metaphoric reflection.
Through generative poetry the mirror `speaks' to the user, each poem unique and tailored to their machine-perceived emotional state.
With this work, we advocate a shift away from `surveillance' style Affective Computing (AC) systems that not only presume, but reinforce, an anachronistic understanding of human emotion, towards user-centred affective interfaces that foster genuine, long-term emotional engagement between people and machines.

\section{Background and Related Work}

\subsection{Computing Emotion}
Affective Computing, first introduced by Rosalind Picard in a 1995 paper \cite{Picard:1997:AC:265013} of the same name, involves the development of computational systems that can recognise, simulate, or express human affect. Picard postulates that in order for computers to be truly intelligent, they, like humans, must possess some level of emotional awareness -- an idea that had already been explored among early AI researchers speculating on the architecture of the human mind. This approach was further supported by more recent advancements in cognitive science that popularised emotional intelligence in contrast to the more traditional measures of human intelligence, such as IQ \cite{Sternberg1981}. Prior to these contributions, emotion was largely perceived to be a mechanism which hindered rational thought, something to be suppressed or ignored. Later research into the neurological basis of emotions \cite{nla.cat-vn1931702} revealed that emotions not only allow for effective communication with others, but they also play a crucial role in high-level decision making and planning.

Affective computing has since found increasing popularity in both research and industry, expanding its applications to include assistive technologies \cite{picard2009future}, mental health monitoring \cite{calvo2016computing,valstar2013avec}, and large-scale surveillance \cite{bullington2005affective}. 
State of the art emotion recognition techniques today attempt to measure all information available in human-human interactions (including facial expression, body language, speech, physiological data), as well as information that is generally inaccessible to humans directly (e.g.~EEG, facial thermal imaging, skin conductivity), to then combine these modes in order to create a more accurate discriminator \cite{ranganathan2016multimodal}.
The vast majority of these techniques are underpinned by a classical theory of human emotion. Under the theory of basic emotion, emotion is divided into a number of emotion categories, of which each is fundamental, universal, and possesses a unique fingerprint that distinguishes it from other emotion categories \cite{ekman1999basic}. 
Although some variant of basic emotion underlies a large majority of emotion recognition research, there has been growing evidence in opposition to the theory within affective neuroscience. 

Over the past two decades, neuroimaging studies have consistently failed to identify localized networks in the brain that correspond to any single discrete emotion category \cite{barrett2013large,Lindquist:2012aa}. These findings, however, conflict with our daily lived experience that is peppered with vivid instances of emotion such as anger, joy, fear, and sadness. The intuitive nature of basic emotion enables it to remain pervasive in both research communities and in wider society.

Neuroscientist Lisa Feldman Barrett provides a solution to this emotion paradox with her Theory of Constructed Emotion \cite{Barrett:2017aa}, in which it is stipulated that discrete emotions are not \textit{natural kinds}, but are instead constructed, in the moment, from a combination of more basic psychological processes.
Specifically, the brain engages in a process of continuous categorisation of interoceptive information, according to available conceptual knowledge, and as informed by a lifetime of embodied experience. Hence, we experience instances of discrete emotion because we have available to us the concepts that allow us to group together, label, and communicate a set of internal and external perceptual information. In the same way we use the concept `blue' to make meaning of 450nm light, we use the concept of `fear' to make meaning of our high heart rate in response to a perceived threat. The TCE holds a number of implications for the field of affective computing

\subsection{Interactive Approaches to Affective Computing}
Adopting the TCE leads to a shift in our understanding of emotion expression and perception within an interaction. In human interaction, the benchmark for emotion perception cannot be accurate prediction, as we have no objective criteria to compare against. Instead, Gendron and Barrett \cite{gendron2018emotion} propose that the benchmark we strive for is the \textit{agreement} between two humans on the meaning of a set of sensory input.
This synchrony between the emotion perceptions of two people is achieved through the iterative process of each one generating and testing their predictions, generally through the use of language. As such, this process can be thought of as the \textit{co-construction} of emotion \cite{gendron2018emotion}. In the same way, the `accuracy' of emotion predictions made by a computational system do not necessarily correspond to the system's understanding of a subjects emotional state, but simply how well the system performs against a set of predetermined criteria (i.e.~predicting the emotion labels in a training set of images).
The system's understanding of a subject's emotion can be determined only by the subject, as they reflect upon the concordance between the systems prediction, and their own internal prediction.

This inference thus motivates a shift away from passive, `surveillance-style' statistical affective computing systems that claim to detect and record human emotion. Emotion is not information that can be measured and transmitted, rather it is a social product that exists only within the frame of an interaction. The call to shift from an informational to an \textit{interactional} approach to AC was first proposed by Boehner et al. \cite{boehner2005affect}, who argue that we move the focus away from the `accurate' measurement of emotion, and instead build systems that allow humans to understand and reflect upon their own emotions in full complexity. The fluid, evolving nature of emotion then requires affective \textit{interfaces} that are dynamic, interactive, and facilitate communication \cite{hook2009affective}.

Interactional approaches to AC typically involve systems where users are encouraged to imbue their own rich interpretation to the presented affective information, rather than being automatically prescribed an emotion label. For example, \textit{Affector} \cite{sengers2005evaluating} is a system developed to communicate emotion and mood between two friends who share an office building. A live video stream of each person is displayed on a screen in the others' office. A visual filter overlays each stream, which is mapped from ambient information collected by sensors, such as movement and temperature. This mapping of each video stream is entirely controlled by the colleague on the other end, allowing them to develop their own unique meaning from the filtered image. In this way, Affector circumvents any need to implement an explicit emotional model. 
Other approaches, such as \textit{Freaky} \cite{Leahu:2014:FPH:2598510.2600879}, utilise statistical-based ER techniques whilst still allowing for open-ended interpretation of the predictions. In these cases, potentially reductionist models of human emotion are implemented, not as an attempt to accurately represent a user's emotion, but simply to facilitate emotional reflection. In this way, we relegate the role of emotion detection technology within these affective interfaces; the machine has no more authority to prescribe your emotional state than another person would.

\subsection{Emotion and Language}
One unique implication of the TCE is the powerful role of language in the experience and perception of emotion. Studies show that it is emotion concepts, as supported by language, that allow for us to classify affect into discrete emotion categories, when there is no physiological or neurological basis that could otherwise allow us to do so \cite{doyle2017language,lindquist2015does}. In one study it was found that once an emotion concept is made inaccessible via semantic satiation, participants have more difficulty in perceiving the emotion category in a pictured facial expression \cite{gendron2012emotion}, suggesting that language can effect how we recognise emotion in others. There is also preliminary evidence that suggests increasing accessibility to emotion concepts via priming can lead us to experience the associated emotion where we otherwise wouldn't have \cite{Lindquist:2008aa}. 
Furthermore, several studies have found that the process of affect labelling (putting our feelings into words) can be seen as a form of implicit emotion regulation, and can lead to a measurable change in physiological markers of affect \cite{torre2018putting,10.1371/journal.pone.0064959}.

These studies illustrate that emotion concepts have a powerful influence over our felt experiences of emotion. Words can not only be used to express emotion, they shape our experience of emotion, and they help to form our perception of emotion in others.  Following this reasoning, it should be possible to utilise language in the development of an affective interface to facilitate a meaningful engagement for users, provoking users to critically reflect on their emotional state and allowing them to engage in a form of emotional regulation. 

The categorisation of affect into one of the basic emotions (happiness, anger, fear, etc) can engage the user in the co-construction of emotion to some extent, however it is extremely limited in scope. Poetic tools such as metaphor and metonym have been studied in their ability to contain and communicate emotion concepts \cite{kovecses2003metaphor}. We turn to poetic text as a vehicle to deliver a large number of diverse and complex emotion concepts, not only enhancing ones ability to express and communicate their emotional state, but expanding on the set of emotional experiences available to them.

\subsection{Mirror Interfaces}
\begin{quote}
    Power and dissatisfaction: for someone who looks at himself can never contemplate himself as pure spectacle. He is at once both subject and object, judge and plaintiff, victim and executioner, torn between what he is and what he knows.
\flushright    --- Sabine Melchior-Bonnet. 1994 \cite{melchior2001mirror}
\end{quote}

From \emph{Narcissus} of Greek Mythology to the magic mirror of \emph{Grimms Fairy Tales}, and their modern incarnations such as Amazon's `Echo Look'\footnote{\url{https://www.amazon.com/Amazon-Echo-Look-Camera-Style-Assistant/dp/B0186JAEWK}}, the mirror is imagined historically and culturally as for more than just a way of literally seeing one's reflection \cite{Pendergrast2003,Melchior-Bonnet2002}. Long established poetic and interactive tropes make the mirror a popular conceptual starting point for many interactive artworks and design concepts, which often subvert or extend culturally familiar concepts drawn from mythology or literature \cite{Bolter2003,Goscilo2010,Albu2016}. 

Artist Daniel Rozin has been building mechanical interactive mirrors since his first `Wooden Mirror' in 1999. These interactive mirrors are distinguished by their material aesthetic and trademark replacement of reflective glass with hundreds of mechanical `pixels', often made from found materials or customised elements that kinetically display the viewers `reflection' sensed by a video camera in real time. However the interactive element of these mirrors is limited to translating the viewer's image into mechanical movement, hence the mirror does not display any autonomy or affective behaviour beyond kinaesthetic translation \cite[Chapter 2]{Bolter2003}. 

Video artist Myron Krueger pioneered many forms of interactive video mirrors since the 1970s. His `Videoplace' \cite{krueger1985}, first presented at CHI'85, was one of the early interactive computer `mirrors' that supported playful interactions between people -- whose silhouettes are projected in front of them -- and a range of virtual interactive objects, such as `Critters' that move over the body like insects moving over a landscape.
Similarly, Camille Utterback and Romy Achituv's `Text Rain' (1999) allowed users' real-time mirrored silhouettes to catch falling letters to form lines of poetry relating to the body and language [\citeNP{Utterback1999}; \citeNP[Chapter 1]{Bolter2003}]. More recently, Rafael Lozano-Hemmer utilized facial recognition technology in his work `Redundant Assembly' (2015), in which multiple users' faces are tracked in real-time and superimposed to create an uncanny composite portrait \cite{Lozano-Hemmer2015}.

Furthermore, \textit{smart mirrors} have become a popular interface element in both research \cite{Rahman2010, Besserer2016, Akpa2017} and commercially. 
In recent years, we have seen smart mirrors extend beyond functional devices, being employed in bespoke, open-ended interfaces that promote ludic engagement \cite{Wallace:2018:SDI:3173574.3173997, Jacobs:2019:PMS:3290605.3300630}. These interactions are designed for momentary, one-off encounters. In contrast, our work aims to move beyond single interactions, generating a longer term engagement that incorporates itself into daily life.

\section{Mirror Ritual} \label{Mirror Design}
\label{s:mirror}

\begin{figure}
\centering
  \includegraphics[width=\columnwidth]{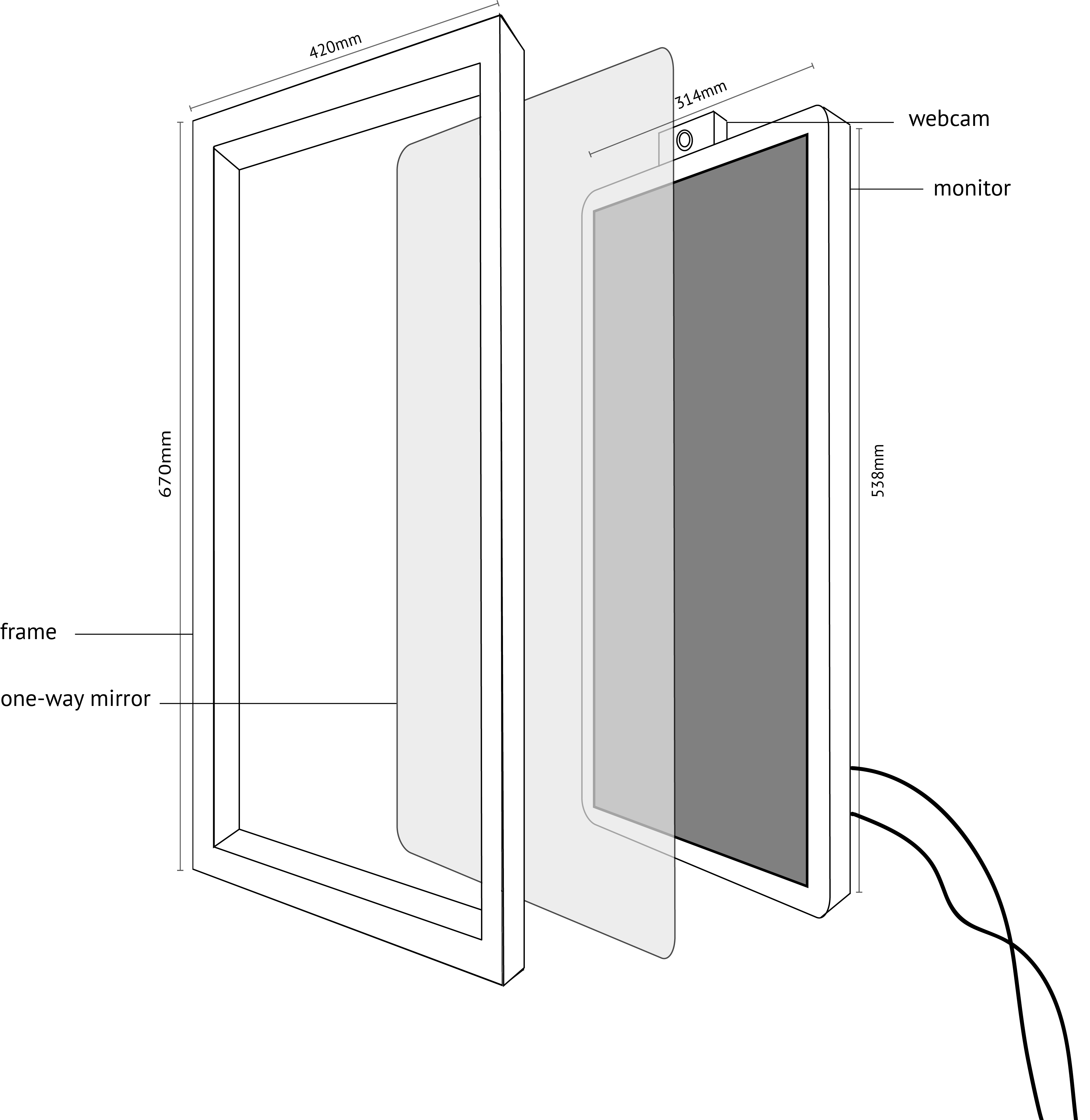}
  \caption{Deconstructed schematic view of the mirror's components: a wooden frame, two-way mirror glass, camera, and display. %\ccbynd~SensiLab, Monash University.
}~\label{fig:schematic}
\end{figure}
We now describe the interface device in detail.
Our device looks and functions like a regular mirror with a wooden frame (400mm x 650mm, see Figure \ref{fig:teaser}). However, hidden behind the mirror glass are a number of technical elements that make it interactive, most importantly a concealed video camera and video display whose image can be seen through the mirror glass (Figure~\ref{fig:schematic}). The monitor itself is not visible, only the image it generates, making the text appear to float on the surface of the glass as is typical with smart mirrors.

The mirror is `activated' when a person, whose face is detected by the system, approaches the mirror and looks at their reflection. As the person looks into the mirror, their current emotion is estimated based on facial expression and this detected emotion is used to generate a unique poem (technical details are provided in the next section). The poem's text gently fades onto the mirror and is displayed for as long as the viewer stares at it.
Turning away from the mirror causes the text to gradually fade away. Looking again at your reflection causes a new poem to be generated based on the currently perceived emotion. The entire process of face detection, emotion classification, and text generation takes approximately 800ms.

\label{ss:mirrorArchitecture}
\begin{figure*}
  \centering
  \includegraphics[width=1.75\columnwidth]{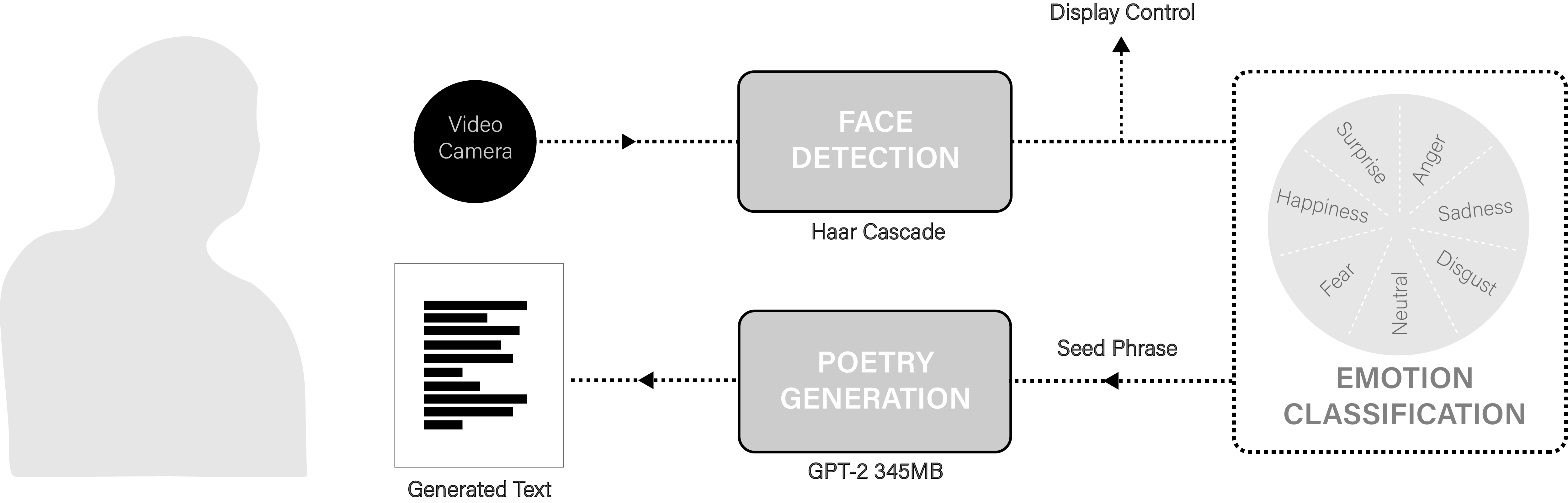}
  \caption{Schematic diagram of the Mirror architecture, showing the flow of information and key stages of data processing and text generation.}~\label{fig:architecture}
\end{figure*}

\subsection{System Architecture} 
Figure~\ref{fig:architecture} shows the main mirror processing architecture. A video camera, concealed behind the mirror, captures real-time video that is then sent to a face detection algorithm (using OpenCV's \cite{opencv_library} Haar Cascades classifier \cite{Viola2004}).
Once a face is detected, the interface performs real-time emotion classification via a Convolution Neural Network (CNN) model trained on the FER-2013 dataset \cite{arriaga2017real} consisting of images of human-annotated facial expressions classified into seven categories: \emph{happiness, sadness, disgust, fear, anger, surprise} and \emph{neutral.}
The real-time classification prediction and associated probability vector is then mapped to an emotion word, which is used as a seed to the text generation model (the \emph{seed text}).

A mapping function, $M: \mathbb{R}^7 \rightarrow S$, where $S$ is a customised set of emotion words, links the confidence of an emotion prediction made by the system to an emotion word of associated strength. For example, a face detected as \textit{happy} with 35\% confidence will return the emotion word `glad', whereas \textit{happy} face detected with 95\% confidence will return the emotion word `ecstatic'. In general, more intense facial expressions correspond to higher confidence, hence confidence is used as an approximate proxy for emotional intensity.
Additionally, the emotion words are preceded by a randomly selected phrase (e.g.~`You are feeling', `You can be') that initiates the generated poem. 

\subsection{Corpus Curation and Training} 
The model being employed for text generation is the GPT-2-345M developed and released by OpenAI \cite{radford2019language}.
The GPT-2 is a general language model with a Transformer based architecture, designed to perform a range of tasks without explicit training. The Transformer architecture was first introduced in \cite{vaswani2017attention}, and has since been used to produce state of the art results in Natural Language Processing (NLP) tasks \cite{DBLP:journals/corr/abs-1810-04805, radford2019language}. Alternative models were considered for this project, however the GPT-2-345M was chosen because it is an open, publicly accessible model, with superior ability to generate coherent text over several paragraphs and relatively easy to customise for bespoke applications. 

For our purposes, we fine-tune the network parameters of the 345M model by retraining it on a custom corpus of selected texts. The choice of training material directly reflects our overall design objectives: to provoke emotional reflection and regulation in the user. We combined a number of different writing styles to construct our corpus, the most prevalent of these being horoscopes \cite{freewill} and postmodern poetry (sourced from \url{poetryfoundation.org}). Texts were chosen to satisfy one of two requirements. Firstly, we intend for the generated text to address the user directly by using the second-person pronouns `you' and `your'. This technique invites the user to make sense of the text through reflection on themselves and their experiences. Users will naturally place themselves as the subject of the mirror's poetry, and through continued use should develop a relationship with the mirror.
The second requirement is that the text involves emotional content. Poetry is a familiar medium through which to communicate complex emotion. We deliberately combine poetry with more informal writing in our corpus in order to generate output that is accessible while remaining abstract and open to interpretation.

Our final corpus is composed of 2,605 individual short-form texts, reaching just over 1MB in size (relatively small for a training corpus). At this size, overfitting our model becomes a concern. We do not want the mirror to simply replicate full poems from our corpus, but to produce original output that blends the various writing styles. We trained our model for 21,000 epochs, reaching an average loss of 0.07, beyond which point we found the model to reproduce exact phrases from the corpus. The text generation process takes as input a number of parameters: \textit{length} is set to 160 for our experiments (upper limit on word length of output); \textit{temperature}, set to 0.8 (higher temperature results in more random completions), and \textit{top\_k} set to 40 (as suggested in the GPT-2 documentation; 40 words are considered at each step of generation). The \emph{seed text} is used to initiate the generation of a poem. Seeding the model with the same seed text does not result in the generation of the same poem.  Finally, the output text is trimmed to prevent run-off sentences, but is otherwise displayed with it's original formatting. This results in some pieces displaying as prose, and others in more traditional poetic format.

\subsection{Design Factors}
As outlined in previous sections, we aim to develop an affective interface that provokes emotional reflection in it's users through the conceptualization of their affective state. These objectives are reflected in the physical design and construction of Mirror Ritual. The use of a mirror surface works symbolically to suggest that users must not only confront their physical reflections, but that they too may be lead to reflect upon their internal emotional state. In addition to the process of reading and interpreting the generated poetry, users are subsequently confronted with their momentary reactions. In this way, Mirror Ritual engages users in the iterative process of the co-construction of their emotional state -- predictions made by the mirror are not intended to be direct representations of a users affective state, however they can work to shape it.

The mirror is developed with a sustained engagement in mind; we intend that users incorporate the Mirror Ritual into their daily routines, developing a meaningful relationship with the interface through multiple encounters over time periods of weeks, months or even years. For this reason, the mirror has been designed to assimilate easily into daily life, both in its aesthetic qualities (i.e.~it appears to be a standard framed mirror), and in it's dual function (it can in most cases simply be used as a standard mirror). In this way, our augmented mirror could replace the conventional mirror that is used in one's daily routine. 
The mirror would ideally be hung in a bathroom, living room, or hallway entrance, creating the space for users to pause and reflect on their mood as they transition between the moments of their day.

\begin{table}[t]
  \centering
  \begin{tabular}{m{6.21cm} m{2cm}}
    % \toprule
    {\small\emph{Emotion Detected}} \\
    {\small \textbf{Seed phrase} and text generated} & {\small Source image}\\
    \midrule
    \multicolumn{2}{l}{\textit{Neutral}} \\
    \textbf{You are relaxed} \newline
    your body is a pillow, \newline
    waiting for me to wake you up \newline
    so you can feel my hands on you \newline
    repeating with ragged breaths \newline
    looking into your eyes \newline
    waiting for you to respond. & \includegraphics[width=2cm]{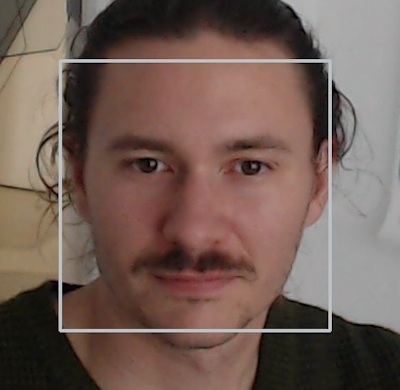}\\
    \midrule
    \multicolumn{2}{l}{\textit{Angry}}\\
    \textbf{So annoyed} you are at the \newline
    thought that your great powers \newline
    might be wasted on irrational fate. \newline
    What have you done \newline
    to deserve such treatment? & \includegraphics[width=2cm]{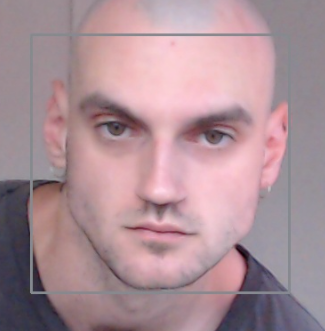}\\
    \midrule
    \multicolumn{2}{l}{\textit{Happy}}\\
    \textbf{You are glad} of the slight \newline
    of body and soul, glad of \newline
    the imperfection of speech, \newline
    and hasten to scatter over \newline
    your soul rays of sunshine, \newline
    and sweet memories, and \newline
    fresh dreams, and bright \newline
    spirits that cut short \newline
    time and space. & \includegraphics[width=2cm]{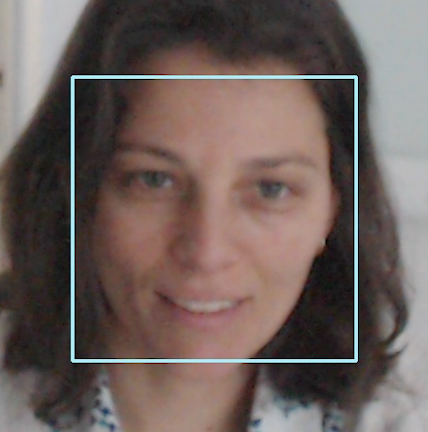}\\
    \midrule
  \end{tabular}
  \caption{Examples of generated poetry. For each poem, the table shows the source facial expression and the subsequent emotion category detected. The strength of emotion categories determines the seed phrase (shown in \textbf{bold}), which is used to generate the poem using a customised GPT-2-345M neural network.} ~\label{tab:poemExamples}
\end{table}

\section{Evaluation}
\label{sec:evaluation}

Our approach to the evaluation of Mirror Ritual is heavily influenced by H{\"o}{\"o}k's proposed two-tier design and evaluation model for affective interfaces \cite{hook2004user}. 
In developing user studies, H{\"o}{\"o}k emphasises the need to allow for the rich interpretation of users' experiences, focusing on individual interpretations rather than attempting to average and summarise results more generally. The overall goal of the evaluation is not to reach objective conclusions, but instead to gain insight about the work's individual effectiveness and further input into the design process.
In this study, we aim to determine whether users can affectively engage with the mirror in a \textit{meaningful} way, with a particular focus on the design choices made.
In future studies, we will determine how, and to what extent the emotion detection component of the interface influences the relationship users develop towards the mirror.

\begin{table}[ht]
\begin{center}
    \begin{tabular}{ r  l }
    \hline
    \textbf{Connectedness} & Connected to the self and the world \\
    \textbf{Coherence} & Making sense of one's experiences \\
    \textbf{Resonance} & Feeling that something is right \\
    \textbf{Purpose} & Sense of core goals, aims, and directions \\
    \textbf{Significance} & Enduring value and importance \\
    \hline
    \end{tabular}
    \caption{The five components of meaning and their definitions \protect\cite{Mekler:2019:FEM:3290605.3300455}. }
    \label{tab:MeaningComponents}
\end{center}
\end{table}

% Recall from the introduction that our overarching aim was to design a system that generated meaningful experiences for its users. 
Recently, Mekler and Hornb{\ae}k presented a framework for the quality of meaning in HCI \cite{Mekler:2019:FEM:3290605.3300455}. Based on a substantive analysis of meaning predominantly from the psychological literature, their framework outlined five components of meaning as a moment-to-moment experience relevant to HCI and UX research: \emph{Connectedness, Purpose, Coherence, Resonance and Significance} (Table \ref{tab:MeaningComponents}). Using this framework as a basis for evaluating meaning, we conducted a study into whether users are able to develop meaningful engagement with the mirror over an extended period of time. A self-assessment questionnaire was developed to evaluate individual responses to each of the five components. The questionnaire consisted of 15 questions (3 for each component) with responses recorded on a 5-point Likert scale. A copy of the questionnaire can be downloaded from 
\url{https://doi.org/10.26180/5d81732551bae}.

In this study, 15 participants (7 male, 8 female) were recruited to undertake a sustained engagement with the mirror, over a period of at least one week. Recruitment was from members of our University through personal contacts, word-of-mouth and flyers advertising the study. No remuneration was provided for participation. Participants' ages ranged from 24 -- 41 with a median of 29. Each participant was asked to incorporate the Mirror Ritual into their daily routine, and to take notes on their experiences. At the conclusion of the study, participants completed the self-assessment questionnaire, along with a semi-structured interview regarding their overall impressions of the mirror. The responses were coded with respect to each of the five components of meaning, a well as our design objectives as outlined in section \ref{Mirror Design}. In this preliminary stage, we chose to approach the evaluation using both quantitative and qualitative methodologies in order to achieve a more comprehensive set of results. We ultimately found the qualitative findings to be the most insightful and constructive for this research. Nevertheless, both quantitative and qualitative findings are reported in this section.

Prior to the study, participants were made aware of the basic functioning of the mirror -- namely that emotion classification is being determined from facial expression, with the result seeding a poetry generation model. For this study, the mirror was placed in our research laboratory, as opposed to in participants' homes.
While this setting is not the ideal environment for the mirror, we compromised on this aspect of the study in order to increase participation numbers which we felt was important for gathering rounded feedback (especially in this initial stage of testing). We did however take a number of steps to ensure the environment was as natural and private as possible.
The mirror was located in a quiet, semi-private space. Participants were given complete freedom to choose the frequency and the length of their interactions, with the only constraint being that they must make at least one visit to the mirror a day.
The study was designed as such to foster natural interactions with the mirror, allowing users the space to freely reflect on their emotional state.

\subsection{Observation of Engagement}
A majority of the participants quickly fell into a comfortable rhythm, working the Mirror Ritual into their existing daily routines. Some opted to make a visit in the mornings before they began their work day, others would stop past before their commute home. The times at which participants chose to interact with the mirror appeared to effect their interpretation of it's messages. Those visiting in the morning would use the poetry to reflect on their goals and attitudes for the day. Participants ending their work day with the mirror tended to frame it's poetry with respect to events that had already occurred. Further discussion into how participants would make meaning of the poetry will be covered in the next section. One participant noted that the mirror felt like an \textit{`everyday companion'} that they would \textit{`miss'} after skipping a visit.
Other participants reported that they visited the mirror only when they had a spare moment, failing to develop a habitual relationship. To some extent, the environmental context of this study prevented participants from seamlessly integrating the mirror into their existing routines -- it was reportedly unusual to gaze into a mirror while in the office. Several participants reported that the presence of a mirror in this context was confronting, noting that they would feel more comfortable interacting with the mirror in a more private setting (e.g.~at home).

\subsection{Meaningful Interaction}
\begin{figure}[htbp]
  \centering
\includegraphics[width=0.45\textwidth]{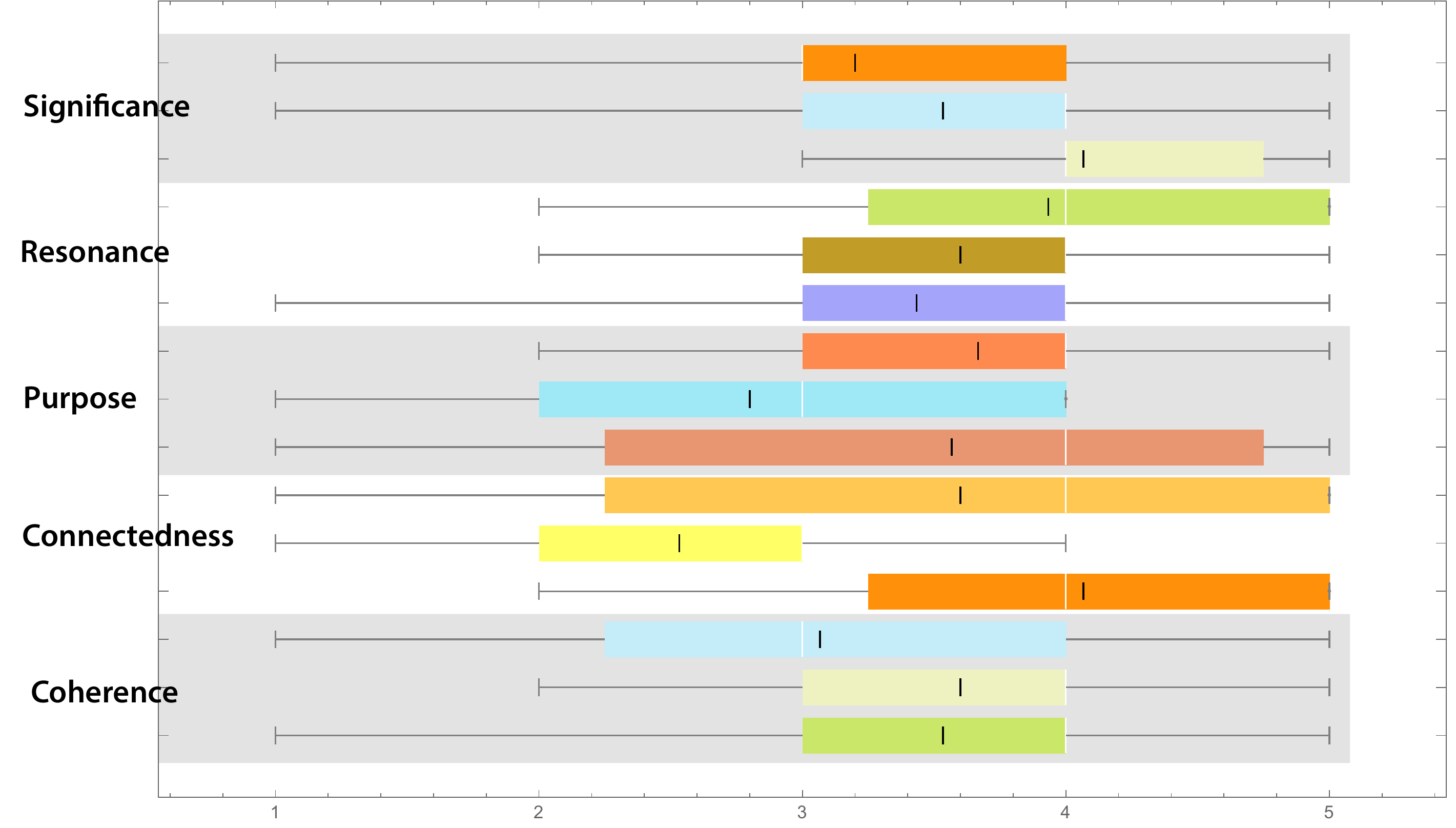}
 \caption{Box and Whisker plot showing the distribution of answers to the self-reported survey questions. Vertical black lines show the mean for each question.} 
 \label{fig:bw}
 \end{figure}

We analysed the responses to the assessment questionnaire for all 15 participants. A box-and-whisker plot (Fig.~\ref{fig:bw}) shows a summary of the responses to each question, with questions grouped by component. Overall the results suggest the interface is able to illicit meaning for its users, but with significant variation between different participants. Responses with the highest means relate to Significance (Q5, $M=4.07,SD=0.70$: \textit{`My Interactions with the mirror were worthwhile'}) and Connectedness (Q1, $M=4.07,SD=0.96$: \textit{`Using the mirror gave me a sense of satisfaction.'}). Only two questions had means with negative outcomes, Q7 (Purpose, $M=2.8,SD=1.01$: \textit{`The mirror helped me to identify new goals strive for'}) and Q6 (Connectedness, $M=2.53,SD=0.92$: \textit{`I felt more connected to the world after interactions with the mirror'}).

We also performed a correlation analysis across each of the individual question responses to see how correlated responses were for each question in the same components and between components. The results showed that, in general questions in the same component category had medium positive correlation. In relation to individual questions, small negative correlations were found between Q8 (Coherence: \textit{`The mirror's poetry was comprehensible to me.'}) and the questions related to Connectedness, Purpose and Resonance, while the positive correlations in the other categories were amongst the weakest for all questions. Similarly, Q9 (Resonance: \textit{`I felt a connection with what the mirror was doing the more I used it.'}) had small negative correlations or very weak positive correlations with the other questions. This supports the findings (discussed below) born out by participant interviews where participants struggled with poems that were too long (Q8) and sought more continuity between experiences (Q9). The strongest correlations ($r=0.81,p<.001$) were between questions related to Purpose (Q12: \textit{` I can envision a use for the mirror in my daily life'}) and Connectedness (Q1, see above).

Although our results give a preliminary indication that participants have, on average, had meaningful interactions with the mirror, we have not yet shed light on \text{how} this process occurs across individual participants. In the evaluation of open-ended, affective interfaces, it is crucial that we understand the single and irreducible experiences of each participant's interactions, framed in the context of their personal emotional landscape. For this reason, we focus on the rich individual interpretations uncovered through semi-structured interviews with participants.
 
\subsubsection{Connectedness}
Many participants mentioned how the displayed poetry led them to feel more connected to the mirror, and to the events going on in their life. One participant reported that the mirror \textit{`was talking directly to me'}, alluding to the mirror's use of the second-person in it's poetry. As discussed in section \ref{Mirror Design}, this style of writing was employed specifically to cultivate a connection with users by addressing them directly. We found that 12 (80\%) of the participants developed some sort of affective relationship with the mirror. In some cases, participants found themselves anthropomorphising the mirror, and ascribing it's messages with intention: one participant proposed that the mirror was aware of their absence on one day, and had mentioned it upon their return, even though each interaction is completely independent in the current version.
Furthermore, we found that a number of participants were able to quickly draw connections from the poetic content, back to their current situation in life.
This experience created a moment for participants to reconnect with themselves, reevaluating and reflecting on their mood, behaviours, or desires.
When participants failed to ground the poetry within their reality, however, they would begin to disengage and withdraw emotionally from the experience.

We furthermore found that the mirror facilitated a connection to others. Participants were able to use particularly illustrative poems as a vehicle through which to communicate their emotional state to others. Since the output is machine-generated, and arguably authorless, there is less stigma attached to the praise or criticism of any one particular poem. Although Mirror Ritual is intended to be a personal and private experience, at least nine participants (60\%) felt compelled to share their poem at one point in the study. We have found that the mirror worked as a catalyst for users to not only reconnect with themselves, but to extend and strengthen connections with others. 

\subsubsection{Coherence}
Coherence describes a user's ability to make sense of an experience, and is of particular interest in the evaluation of machine-generated poetry. 
We found that all of the participants would naturally engage in the sense-making process when first presented with a poem. Many reported that this process was often challenging and required reflexive thought. At times, participants felt that the presented poem was \textit{`too long'}, such that it became nonsensical. On the other hand, participants were understandably disappointed when the mirror output only a few words (e.g.~`You are frustrated'). There appears to be a sweet spot in the ideal length of a given poem. When the poem is too short, the user finds meaning too easily, and is not challenged to reflect more deeply. Too long, and the user becomes lost in the multitude of conflicting interpretations.

A number of participants found the sense-making process itself to be the most rewarding aspect of the experience. One participant reported that the poetry \textit{`Didn't necessarily make sense, but it made sense with me'}. Most participants were prepared to spend time interpreting the poetry and making it \textit{`fit'}, with five participants (33\%) reporting this exercise to be cathartic. For nine participants (60\%), the mirror's message became the lens through which they could decode the complex emotions surrounding their recent experiences. In one example, a participant had a particularly difficult day in which a student confided in them about a traumatic experience. When the participant visited the mirror later in the evening, they received the following message: 
\begin{quotation}
\noindent You are relaxed and as free as can be\\
in a cage made up of rocks\\
and I am still in here, chained to the cage\\
with lots of holes to sit in and no freedom at all
\end{quotation}
They reported that this poem was an apt expression of how they were feeling about the situation.
In the case of Mirror Ritual, the ability for participants to comprehend their experience overlaps heavily with connectedness; participants make sense of the poetry by connecting it back to the events in their life. We found that these two components of meaning would tend to co-occur due to the personal nature of the themes raised by the mirror. In some cases, participants were able to make sense of the poetry, but could not personally connect with it. Even when opposing the mirror's suggestions, however, participants were still required to reflect on their mood to some extent. Yet when the experience was both nonsensical, and disconnected from their greater reality, at least five (33\%) participants report disengaging from the interaction and withdrawing emotionally. Due to the probabilistic nature of our language generation model, it is difficult to control for this variation in the coherence of the output.

\begin{figure*}[t]
  \centering
\includegraphics[width=0.95\textwidth]{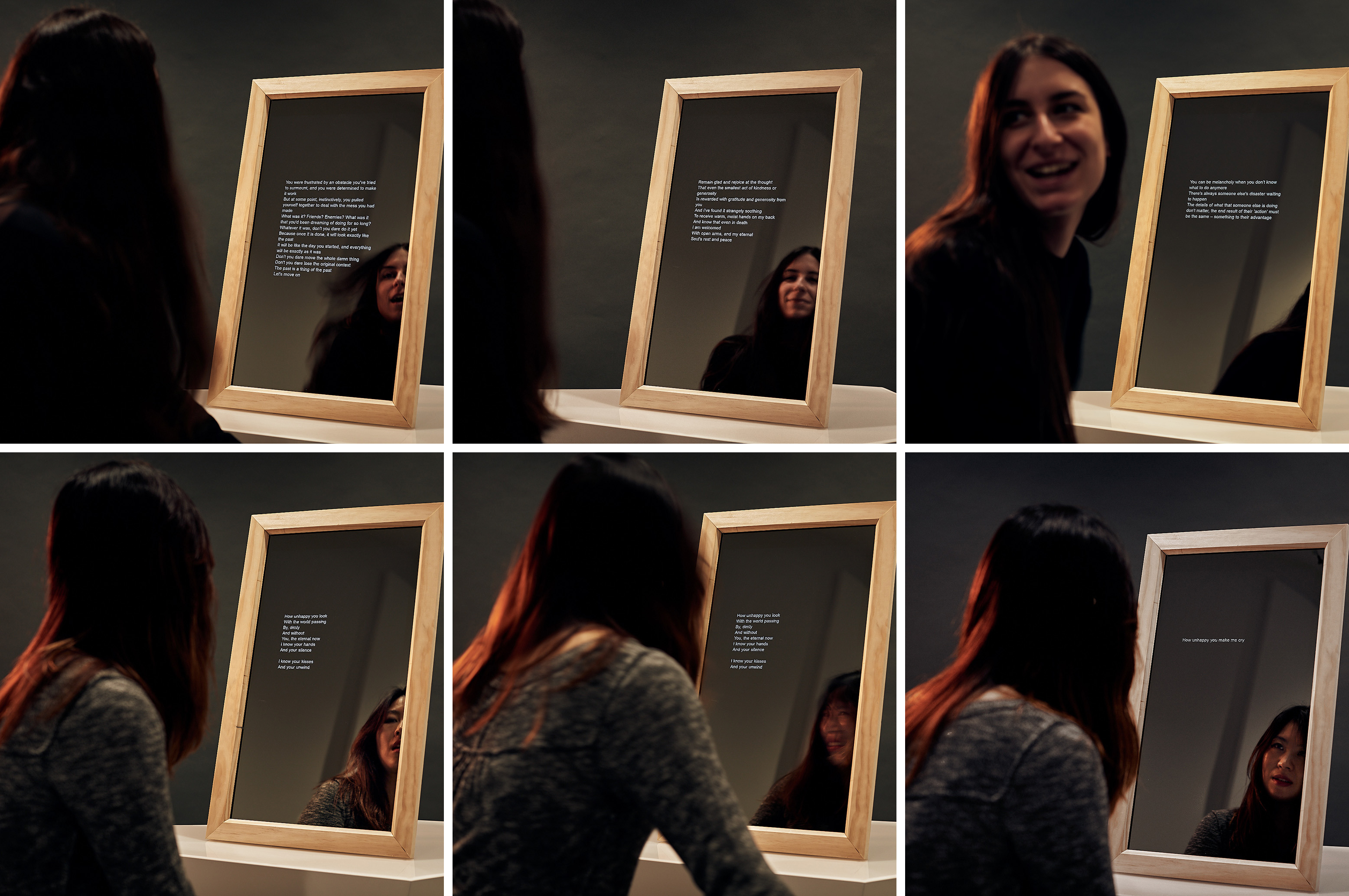}
 \caption{Sequence of interactions with the mirror in the evaluation environment.} 
 \label{fig:double}
 \end{figure*}
 
\subsubsection{Resonance}
As explained in \cite{Mekler:2019:FEM:3290605.3300455}, resonance differs from coherence in that it does not require active reflection, but instead describes the visceral and immediate `feeling'. In designing an interface that will affectively engage users, resonance is perhaps the most vital component. The mirror's ability to resonate with users was seen most clearly in moments where it caused an unexpected, but pleasant, disruption. Two participants reported sitting around the mirror, engaged in a conversation about whether love necessarily leads to attachment and loss. At this point, the following poem appeared on the mirror:
\begin{quotation}
\noindent So melancholy\\
or sad as I look out at the dark\\
is the stream\\
it seems to me I'm going to fall in love\\
it seems\\
I think I'm in love\\
it seems
\end{quotation} The participants paused their conversation to pay attention to the mirror, noting the uncanny relevance of the poem. From the outside, it may be difficult to understand what exactly the poem stirred up in either participant. Yet, the impulse to interrupt the conversation occurred instantaneously, before either participant had the opportunity to actively interpret the poetry.

At least 12 participants (80\%) described moments where the displayed poetry instantly triggered an emotional response, before any meaning could be imbued. Ten participants (67\%) reported laughing at the absurd nature of some of the poems. In one example, a participant reports having experienced a remarkably unlucky string of events happen to them throughout the day, which culminated with them falling of their bike on the short ride home from work. The next morning, they received the following message: \begin{quotation}
\noindent You should feel elated and uplifted and reminded of how blessed you are for having the discipline and the resilience necessary to navigate the many varied situations that so many others find it hard to deal with.
\end{quotation}
Despite it coming from a machine, this consoling message reportedly brought the participant to tears. Strong affective reactions like these were triggered only in a small number of interactions overall. Yet it was these interactions that left the deepest impression on participants, heavily shaping the perceived meaningfulness of Mirror Ritual as a whole.

\subsubsection{Purpose}
In this study, participants were not given any particular task or goal in their interactions. The reasons for interacting with the mirror were left for the participants themselves to discover. This was done deliberately such that we could gain more insight into \textit{why} participants choose to interact with the mirror. 
Seven participants (47\%) described using the mirror out of curiosity; either trying to ascertain how it works, or simply interested in how the mirror perceives them.
The motivation for these interactions seems to derive from the element of novelty.
Another five participants (33\%) reported that over time, they began to use the mirror as an excuse to pause and reflect, describing the experience as \textit{`meditative'}, \textit{`therapeutic'}, and \textit{`mindful'}. Among these participants, several reported that they can envision a use for the mirror.
Only three participants (20\%) failed to develop any purpose for the mirror that is transferable to their daily life.

A number of participants noted that the experience lacked continuity across visits. The variability in the mirror's output left participants feeling that their experiences were disjointed, and struggling to summarise their interactions in a few statements.
We found that although Mirror Ritual generated overall positive, constructive experiences for participants in the moment, they failed to develop clear motivations to use the mirror long-term.

\subsubsection{Significance}
Significance is arguably the most difficult component to design for. The interaction should not only provide momentary satisfaction, but should also create lasting value for users. We conducted a follow-up interview, three weeks following the conclusion of the study, in an effort to determine if and how Mirror Ritual left an enduring impression on participants.

A couple of participants reported that interactions with the mirror led them to be more conscious of how they express emotion, causing them to be more self-aware in general. One participant explained that their overall experience left them with a greater appreciation of \textit{`momentary encounters'}. Another two (13\%) participants felt that the mirror's suggestions helped them to explore different perspectives on their current circumstances. The perspectives offered by the mirror were not necessarily original, but instead served as a reminder of what the participants already knew. In this way, the mirror simply \textit{`put into words'} the feelings that participants had been experiencing. Nine participants (60\%) were compelled to take a photo with at least one particularly affecting poem, with many reporting that they had revisited the poem at a later time. One participant noted that experience extended beyond momentary interaction, as the generated poem is something that you \textit{`take with you'}.

\section{Discussion}
Our evaluation of Mirror Ritual has revealed that participants naturally engage affectively with the mirror -- although the feelings elicited varied significantly across individuals. Our use of the survey questions was an initial attempt to devise a statistical instrument for evaluating meaning and its components as outlined in \cite{Mekler:2019:FEM:3290605.3300455}. This instrument needs further refinement and validation before drawing any significant conclusions on its effectiveness or its results. The relatively small sample size and use of convenience sampling for our survey makes any broad generalisations difficult. Nonetheless, the initial results showed moderate to high correlation between some questions in each category, which supports further refinement of the questions and evaluation with larger sample sizes. Generally, similar survey instruments take many years and numerous revisions before becoming accepted as reliable (e.g.~\cite{Jackson2010}).

The qualitative analysis revealed a number of important findings that feed directly into future design revisions, along with valuable information on how and why some participants found the mirror meaningful. The open-ended nature of the poetry led participants to \textit{`fill the gaps'} by imbuing their own personal meaning to the mirror's messages. The mirror appeared to bring to the surface dormant emotions of participants by allowing them to conceptualize, and in some cases verbalize, what they have been feeling.

The custom language model plays an important role in generating these responses from participants -- our corpus was specially curated to elicit self reflection of this nature. Even so, what the text generator was able to produce surprised and excited us as curators of the training corpus. Additionally, the end-to-end interaction, including the physical design and interaction design, contribute significantly to the meaningfulness of these experiences. Many participants enjoyed that each poem was entirely unique, often taking ownership of the poem, and feeling that it is a personalized message made specifically for them, and for that moment in time. 
% We found that one's conception of the mirror and how it functions, heavily shapes their reaction to it's poetry. Despite the emotion recognition model crude, and often misreading 
% We found that it is difficult to attribute the meaningfulness of the experience to any one component of the system.
We found that the most meaningful interactions occurred when participants were able to come to their interpretations easily, especially when these interpretations were concerning their current circumstances. In these cases, many report feeling that the mirror would say \textit{`the right thing at the right time'}. Overall, we found that 11 (73\%) participants were able to forge a meaningful relationship with the mirror.

\section{Future Work}

We identified a number of factors that caused participants to emotionally detach from the experience. Below we list our findings and features to be addressed in the next iterations of Mirror Ritual:
\subsubsection{Poetry Generation}
Almost all participants reported that they found the mirror's poetry to be incoherent at times --- especially in the case of longer poems. Currently, word length is passed in as a parameter to the generation system, and is set at 160. We found that reducing this limit to 80 eliminated the majority of lengthy poems, and hence should increase the perceived coherence of a user's experience overall. Additionally, participants found it difficult to engage with exceptionally short poems. A validation stage can be added that rejects any poems shorter than five words (in which case the system simply regenerates a new poem using the same seed phrase). 

\subsubsection{Interaction Design}
It was found that many participants struggled to develop a direction in their interactions with the mirror. Our study discovered that users require some overarching narrative in the experience, one that links their past interactions, and gives impetus for future interactions. In the next iterations of our design, we will to add `memory' into the system. For example, the mirror could track the moods of each user over time, and generate poetry that reflects these patterns. Ideally, the experience would extend beyond momentary insights, and instead be actively utilised as a tool for ongoing emotional reflection and regulation. 
Furthermore, the mirror could also track a user's emotion before, during, and after the reading of a poem. This information can be used to further personalize generated poetry towards individual users, by recording which poems triggered strong affective responses, and over time constructing individual user `personalities'.

\subsubsection{Experimental Design}
Several participants noted that engaging with the mirror in their workplace felt `confronting'. Additionally, we found that participants impressions of the mirror transformed drastically over the week long study. Nine participants (60\%) reported that their idea of the mirror had changed significantly since their first use, with one participant remarking \textit{`the more I used it, the more relevant it felt'}. These findings will allow us to further refine the experimental set-up. Namely, the mirror will be set-up in participants homes, fostering more personal, private interactions that seamlessly integrate into daily life. Furthermore, the length of the study will be extended from one week, to span several weeks, allowing us to better determine how Mirror Ritual can transform one's lived emotional experiences.

\section{Conclusion}
\begin{quote}
    If we only look \emph{through} the interface,\\
    we cannot appreciate the ways in which\\
    the interface itself shapes our experience.
    \flushright --- Bolter \& Gromala \cite{Bolter2003}
\end{quote}

Mirror Ritual is a bespoke affective interface designed to bridge the gap between two seemingly incongruous areas of research: automatic emotion classification as developed in affective computing, and constructed emotion as founded in affective neuroscience. 
We do this by relegating the role of AI from one of objective measurement to that of subjective perception. 
We utilize existing emotion classification techniques, not as an instrument to measure one's `true' emotional state, but to instead engage the user in the dynamic and iterative conceptualization of their feelings and experiences.

Affectively-charged, machine generated poetry provides users with rich and unique conceptualisations that -- through the process of critical self-reflection -- they can choose to affirm or resist.
We presented our findings in a preliminary investigation into the mirror's ability to foster meaningful experiences. Mirror Ritual is able to emotionally engage users in momentary encounters, as well as forging a sustained and evolving affective relationship with the majority of participants. A qualitative analysis illustrated the process through which users make meaning of the AI generated poetry, using their personal lived experiences to frame the mirror's messages.

In addition to practically applying the theory of constructed emotion to affective interfaces, the experimental design and findings presented here make up a significant portion of the overall research. Through experimentation, we identified a number of design changes, but more importantly we presented a detailed exploration into how participants can engage in the co-construction of emotion, making a case for the development of user-centered affective interfaces over surveillance style emotion detection technology. 

Furthermore, our study showed that the objective evaluation of such open-ended interfaces is difficult. Formalized evaluations appear less effective in identifying the design considerations that influence an individual's experience with affective interfaces \cite{hook2004user}. Our study is useful to the CHI community by highlighting both the opportunities and drawbacks of this style of research. We have demonstrated how such affective interfaces, designed purely to promote awareness and reflection, can be useful and overall, beneficial to users. We believe that this study provides impetus for future investigation into the idea of human-machine co-construction of emotion, which is a novel and still somewhat controversial idea in HCI research. 

\balance{}

% REFERENCES FORMAT
% References must be the same font size as other body text.
\bibliographystyle{SIGCHI-Reference-Format}
% \bibliography{references,affective,MyReferences}
 \bibliography{extracted}
\end{document}